\documentclass{caosp310}

\usepackage{natbib}
\usepackage{enumitem}
\usepackage{orcidlink}
\usepackage{amsmath}

\articleNo{361}
\pubyear{2024}
\volume{54}
\volnumber{3}
\firstpage{1}
\received{XX XX, XXXX}
\accepted{XX XX, XXXX}

\def\BibTeX{{\rm B\kern-.05em{\sc i\kern-.025em b}\kern-.08em
             T\kern-.1667em\lower.7ex\hbox{E}\kern-.125emX}}

\begin{document}

%
\hauthor{W. H.\,Elsanhoury}

\title{Kinematics of High-Velocity Stars Utilizing LAMOST and Gaia DR3 Archives within 100 kpc}


%
%
\author{
        W. H.\,Elsanhoury$^{1,2}$~\orcid{0000-0002-2298-4026}}

%
\institute{$^1$Physics Department, College of Science, Northern Border University, Arar, Saudi Arabia.\\
$^2$Astronomy Department, National Research Institute of Astronomy and Geophysics (NRIAG), 11421 Helwan. Cairo, Egypt.\\ \email{elsanhoury@nriag.sci.eg $\&$ elsanhoury@nbu.edu.sa}}


\date{March 8, 2003}

\maketitle

\begin{abstract}
The kinematic parameters identified from high-velocity stars situated within $\sim$ 100 kpc are examined and analyzed. We included three high-velocity programs comprising 591, 87, and 519 stars as a function of distances ranging from 0.10 to nearly 109 kpc. In this analysis, we will determine the spatial velocities ($U,~V,~W$) in galactic coordinates along with their velocity dispersion ($\sigma_1,~\sigma_2,~\sigma_3$), the convergent point ($A_o,~D_o$), and therefore, the solar motion ($S_{\odot}$). In conclusion, we are calculating the first Oort constant ($A=11.94 \pm 0.29$ km s$^{-1}$ kpc$^{-1}$) and the second one (i.e., $B=-17.78 \pm 0.24$ km s$^{-1}$ kpc$^{-1}$), the angular rotation rate $|A – B| = 25.07 \pm 5.01$ km s$^{-1}$ kpc$^{-1}$, and the average rotational velocity $V_o = 243.72 \pm 15.61$ km s$^{-1}$.

\end{abstract}
%
\section{Introduction}
\label{intr}
The stars known as high-velocity stars (HiVels) travel through space at substantially faster rates than the average Milky Way (MW) Galaxy star. These objects are special in astronomy study since they can reach speeds of thousands or even hundreds of kilometers per second. In the disk region, almost all the stars in our Galaxy rotate around the galactic center at a normal velocity of 200–240 km s$^{-1}$ \citep{2012ApJ...759..131B,2016MNRAS.463.2623H,2019ApJ...871..120E}. An alternative indirect estimate of both the circular velocity ($V_o$) and the escape velocity ($V_{esc}$) at which the stars in the solar neighborhood would have sufficient energy to completely escape from our Galaxy's gravitational field can be obtained from the kinematic properties of the halo-population stars that have been observed to have the largest space velocities to the Sun (i.e., the extreme HiVels).

Including the Gaia DR3, Large Sky Area Multi-Object Fiber Telescope (LAMOST, \cite{2012RAA....12.1197C}) galactic surveys, and spectroscopic observations from large-scale galactic surveys (SEGUE, \citep{2009AJ....137.4377Y,2022ApJS..259...60R}) have demonstrated the existence of HiVels in our Galaxy. Among these, a few are even hypervelocity stars (HVSs), meaning that their $V_{esc}$ is less than their overall galactocentric velocities ($V_{GSR}$), and almost all HiVels have low luminosity ($M_G \sim 10$ mag).

In the halo, they exhibit tens of kilometers per second \citep{2008ApJ...684.1143X,2016MNRAS.463.2623H}, especially when a star approaches or even exceeds the Galaxy's escape velocity at its position. HiVels indicate the presence of extreme dynamical and astrophysical processes \citep{1988Natur.331..687H,2003ApJ...599.1129Y,2006ApJ...653.1194B,2009ApJ...691L..63A,2008MNRAS.383...86O,2015MNRAS.454.2677C,2019MNRAS.490..157M}. The finding of such rare objects offers a valuable tool for investigating the MW's mass distribution, particularly its dark component \citep{2005ApJ...634..344G,2017MNRAS.467.1844R,2019MNRAS.487.4025C}, because they travel large distances across it \citep{2005ApJ...634..344G,2008ApJ...680..312K}, and their trajectories can also be used to probe the shape of the Galaxy’s dark matter halo \citep{2006ApJ...653.1194B,2007MNRAS.379.1293Y}. The study of stellar motion and the dynamics of the MW Galaxy reveals a relationship between HiVels and the Oort constants ($A~\&~B$), where HiVels offer special test cases for comprehending extreme stellar motions both inside and outside of the Galaxy.

HiVels and HVSs can be divided into four subclasses, each with a distinct origin of these high velocities; i) black hole ejection (BHE): As a result of tidal interaction between a close stellar binary system and a supermassive black hole (SMBH) in the Galaxy, a process known as the "Hills mechanism," the so-called HVSs (with velocities even greater than 1000 km s$^{-1}$) were first predicted from theoretical arguments of \cite{1988Natur.331..687H}. Extending the Hills mechanism allows for the ejection of HVSs and HiVels, ii) supernova explosions (SNEs), can cause significant disruption to binary systems and induce their companion stars to become HiVels or HVSs. Examples of such explosions include core-collapse and thermonuclear supernova explosions (SNe) \citep{1961BAN....15..265B,2000ApJ...544..437P,2009A&A...493.1081J,2009A&A...508L..27W,2013ApJ...770L...8P,2013ApJ...771..118Z,2018ApJ...865...15S,2020A&A...641A..52N,2019ApJ...887...68B}. Generally, stars with velocities of no more than 300–400 km s$^{-1}$ cannot be ejected by core-collapse SNEs \citep{2000ApJ...544..437P}, iii) dynamic ejection mechanism (DEM) as proposed by \cite{1967BOTT....4...86P}, into which a possible explanation for the runaway stars in the galactic Outer Belt. This process involves the expulsion of runaway stars from young stellar clusters because of close stellar interactions. This mechanism often achieves a maximum kick velocity of around 300–400 km s$^{-1}$, which is the result of collisions between two close binaries \citep{1990AJ.....99..608L,1991AJ....101..562L,2009MNRAS.396..570G}, and iv) tidal stripping from dwarf galaxies (TSD): In this scenario, stars can be rapidly removed from a dwarf Galaxy that is pericentrically passing through a region where the MW gravity field is causing tidal disruption \cite{2009ApJ...691L..63A}. According to \cite{2011A&A...535A..70P}, runway stars in this mechanism must be expelled by a large dwarf Galaxy ($>10^{10} M_\odot$).

The study aims to examine and report the spatial structures, kinematics of the HiVels, including their velocity ellipsoid motion characteristics, parameters characterizing the local rotational properties of our Galaxy such as Oort’s constants $A$ and $B$. In the context of our ongoing investigations into stellar associations, we present velocity ellipsoid parameters for three Program stars as a function of distances ($d$), i.e., Program I (591 stars; $0.10 \le d$(kpc)$~\le 15.40$), Program II (87 stars; $0.30 \le d$(kpc)$~\le 108.64$), and Program III (519 stars; $0.29\le d$(kpc) $\le 16.44$). Moreover, determination of the equatorial coordinates for convergent points (i.e., $A_o,~D_o$) with AD-chart method. Finally, we computed the MW Galaxy’s local differential rotation close to the Sun, i.e., the Oort constants ($A~\&~B$) based on observed velocities of three considered Program stars.

The remainder of this article is structured as follows: Section \ref{sec2} provides the selected data considered in this analysis. Section \ref{sec3} details the computational methods, including velocity ellipsoid parameters and the convergent point Section \ref{sec4} deals with the galactic rotational constants. We close finally with discussion and conclusions in Section \ref{sec5}.

\section{Selected data}
\label{sec2}

Large Sky Area Multi-Object Fiber Spectroscopic Telescope (LAMOST), also named the “Guo Shou Jing” Telescope \citep{2012RAA....12.1197C} is a 4-m Schmidt spectroscopic survey telescope specifically developed to study four thousand targets per exposure in a field of view roughly 5$^{o}$ in diameter \citep{2012RAA....12.1197C,2012RAA....12.1243L,2014IAUS..298.....F}. LAMOST spectra have a resolution of $\sim$1800 with a wavelength range of 3800–9100 $\AA$. In March 2020, 10,608,416 spectra in DR7\footnote{\url{http://dr7.lamost.org/}} \citep{2020ApJ...889..117L} were made accessible by LAMOST.

The third major data release from the European Space Agency's (ESA) Gaia mission, which aspires to provide the most precise three-dimensional map of the Milky Way, is called Gaia Data Release 3 (hereafter DR3) \citep{2023A&A...674A...1G}. Gaia is an astrometry expedition that was launched in 2013 to determine the locations, distances, movements, and other characteristics of over one billion stars and other celestial objects. Released on June 13, 2022, DR3 represents a major update to the earlier Gaia data releases (DR1 and DR2).

Understanding stars' physical characteristics and chemical makeup. Moreover, detailed astrometric parameters like equatorial coordinate system ($\alpha$, $\delta$), parallaxes ($\pi$; mas), movements within the MW as well-known the proper motion (PM; mas yr$^{-1}$) components in right ascension and declination ($\mu_{\alpha}\cos{\delta}$, $\mu_{\delta}$) for roughly 470 million stars, and radial velocities ($V_r$; km s$^{-1}$) for 34 million stars, DR3 is a rich source of stellar data and provides photometric across three broadband filters: the $G$ band (330–1050 nm), the Blue Prism ($G_{BP}$: 330–680 nm), and the Red Prism ($G_{RP}$: 630–1050 nm) for sources brighter than 21 mag. Common errors in the photometric observations throughout these three bands with $G$ $\le$ 20 magnitudes are approximately 0.30 magnitudes and are increased for fainter stars (approaching $G=$ 21). DR3 can measure $PM$ for the bright stars ($G$ $\le$ 15) with remarkable accuracy, frequently to within $\sim$ 0.02 to 0.03 mas yr$^{-1}$. With uncertainties of about 1.00 mas yr$^{-1}$ or exceed for fainter stars ($G=$ 20), but still yields useful motion data. The uncertainty limit in parallax is about 0.02 to 0.03 mas for $G$ $<$ 15 mag, $\sim$ 0.07 mas for $G=$ 17 mag, $\sim$ 0.50 mas for $G=$ 20 mag, and $\sim$ 1.30 mas for $G=$ 21 \citep{2023A&A...674A...1G}. The astrometric accuracy of DR3 is significantly better than that of DR2, with $PM$ accuracy being doubled and parallax accuracy being roughly 1.5 times higher. Furthermore, astrometric inaccuracies in parallax measurements were reduced by 30-40$\%$, while accurate motion measurements were improved by 2.5 times.

\cite{2021ApJS..252....3L} reported on approximately 591 HiVels in the galactic halos with three-dimensional velocities in the galactic rest frame larger than 445 km s$^{-1}$ that were chosen from over 10 million spectra of Data Release 7 of the Large Sky Area Multi-object Fiber Spectroscopic Telescope (LAMOST DR7) and the second Gaia data release (Gaia DR2, \cite{2016A&A...595A...1G}).

According to the fifth data release of the $V_r$ Experiment survey (RAVE DR5, \citep{2017AJ....153...75K}), the twelfth data release of the Sloan Digital Sky Survey (SDSS DR12, \citep{2015ApJS..219...12A}), the eighth data release of Large Sky Area Multi-Object Fiber Telescope (LAMOST DR8, \citep{2022ApJS..259...51W}), sixteenth data release of The Apache Point Observatory Galactic Evolution Experiment (APOGEE DR16, \citep{2017AJ....154...94M}), second data release of the Galactic Archaeology (GALAH DR2, \citep{2015MNRAS.449.2604D}), and Early Data Release 3 (Gaia EDR3), \cite{2023AJ....166...12L} presents about 88 HiVels by large-scale galactic surveys.

With crossmatch between DR3 and precise $V_r$ with other large-scale galactic surveys, such as GALAH DR3 \citep{2015MNRAS.449.2604D,2021MNRAS.506..150B}, LAMOST DR10 \citep{2009AJ....137.4377Y,2012RAA....12..723Z}, RAVE DR6 \citep{2020AJ....160...83S}, APOGEE DR17 \citep{2017AJ....154...94M}, \cite{2024AJ....167...76L} presents about 591 HiVels of the Large Magellanic Cloud (LMC).

In what follows, we attention to bold here and report in our analysis three Programs (I, II, and III) of halo HiVels as a function of distances and lags between a few parsecs till $\sim$ 100 kpc.

The most recent data from DR3 was used to update and enhance Programs I, II, and III. To ensure data consistency crossmatches were carried out using software that was based on the Tool for OPerations on Catalogues And Tables (TOPCAT) and Starlink Tables Infrastructure Library (STIL; \cite{2005ASPC..347...29T}). This tool has several options for modifying astronomical catalogs and is especially reliable when examining tabular data within a given range ($0 < x < 1$).

\begin{enumerate}[label=(\roman*)]
\item The Program I mainly aim to comprehend the galactic and kinematic characteristics of about 591 HiVels \citep{2021ApJS..252....3L} with LAMOST and Gaia DR3 located ($0.10\le d$(kpc) $\le 15.40$).
\item Program II was set for 87 HiVels who are ejected from the galactic center with large-scale galactic surveys \citep{2023AJ....166...12L} with DR3, LAMOST, SDSS, and other large-scale surveys combining astrometric and spectroscopic data within ($0.30\le d$(kpc) $\le 108.64$). The missed one of 88 stars is HVS23 ($\alpha = 240^o.537580$, $\delta = 0^o.912272$) into which there is no Gaia DR3 identification number and no proper motions data in both directions, therefore, we neglect it from this Program set.  
\item Program III was set and updated via DR3 sources for 519 HiVels influenced by LMC’s gravitational potential \cite{2024AJ....167...76L} in the range of ($0.29\le d$(kpc) $\le 16.44$).
\end{enumerate}

Table \ref{tab1} presents the fundamental parameters of these three Programs of HiVels \citep{2021ApJS..252....3L,2023AJ....166...12L,2024AJ....167...76L}, respectively. The distribution of selected stars across the sky is shown in Figure \ref{fig1} with almost random directions of velocity vectors. Figure \ref{fig2} presents the $V_r$ distribution as a function of galactic longitude ($l^o$) for all halo HiVels observed in our considered three Programs (I, II, and III).

\begin{table}[t]
\label{tab1}
\small
\begin{center}
\caption{The fundamental parameters of the three Programs I, II, and III of HiVels adopted by \cite{2021ApJS..252....3L,2023AJ....166...12L}, and \cite{2024AJ....167...76L}, respectively.}
\label{t1}
\begin{tabular}{lllll}
\hline\hline
No.  & Gaia DR3 ID &   Ra. (deg.) & Dec. (deg.) & $V_r \pm \sigma_{V_r}$ (Km s$^{-1}$)\\
\hline
Program I: $N$ = 591 stars&&&&\\
$0.10 \le d$(kpc)$~\le 15.40$&&&&\\
\hline
1&1383279090527227264&240.3375&41.1668&-184 $\pm$ 5.00\\
2&1570348658847157888&193.4372&55.0581&-230 $\pm$ 15.00\\ 
3&966594450238136704&102.4834&46.8368&-80 $\pm$ 6.00\\
. & . & . & . & .\\
. & . & . & . & .\\
. & . & . & . & .\\
591&1272009269712167680&229.6719&28.3365&-147 $\pm$12.00\\
\hline
Program II: $N$ = 87 stars&&&&\\
$0.30 \le d$(kpc)$~\le 108.64$&&&&\\
\hline
1&255667999133782348&12.3624&7.1290&-345.37 $\pm$ 9.48\\
2&2801887851883799936&12.5096&21.4207&-32.94 $\pm$ 14.38\\
3&2510946771548268160&29.3032&1.1937&217 $\pm$ 2.26\\
. & . & . & . & .\\
. & . & . & . & .\\
. & . & . & . & .\\
87&2872564390598678016&352.2706&33.0032&237.3 $\pm$ 6.40\\
\hline
Program III: $N$ = 519 stars&&&&\\
$0.29 \le d$(kpc)$~\le 16.44$&&&&\\
\hline
1&6087590373666222080&201.7684&-44.5009&328.84 $\pm$ 1.19\\
2&6098873935647575552&220.6412&-44.5675&163.15 $\pm$ 0.71\\
3&6135358205355639424&193.4286&-44.9466&203.66 $\pm$ 0.32\\
. & . & . & . & .\\
. & . & . & . & .\\
. & . & . & . & .\\
519&4421621689671197568&229.5736&2.1771&-54.16 $\pm$ 0.82\\
\hline\hline

\label{tab1}
\end{tabular}
\end{center}
\end{table}

\begin{figure}
\label{fig1}
\centerline{\includegraphics[width=0.75\textwidth,clip=]{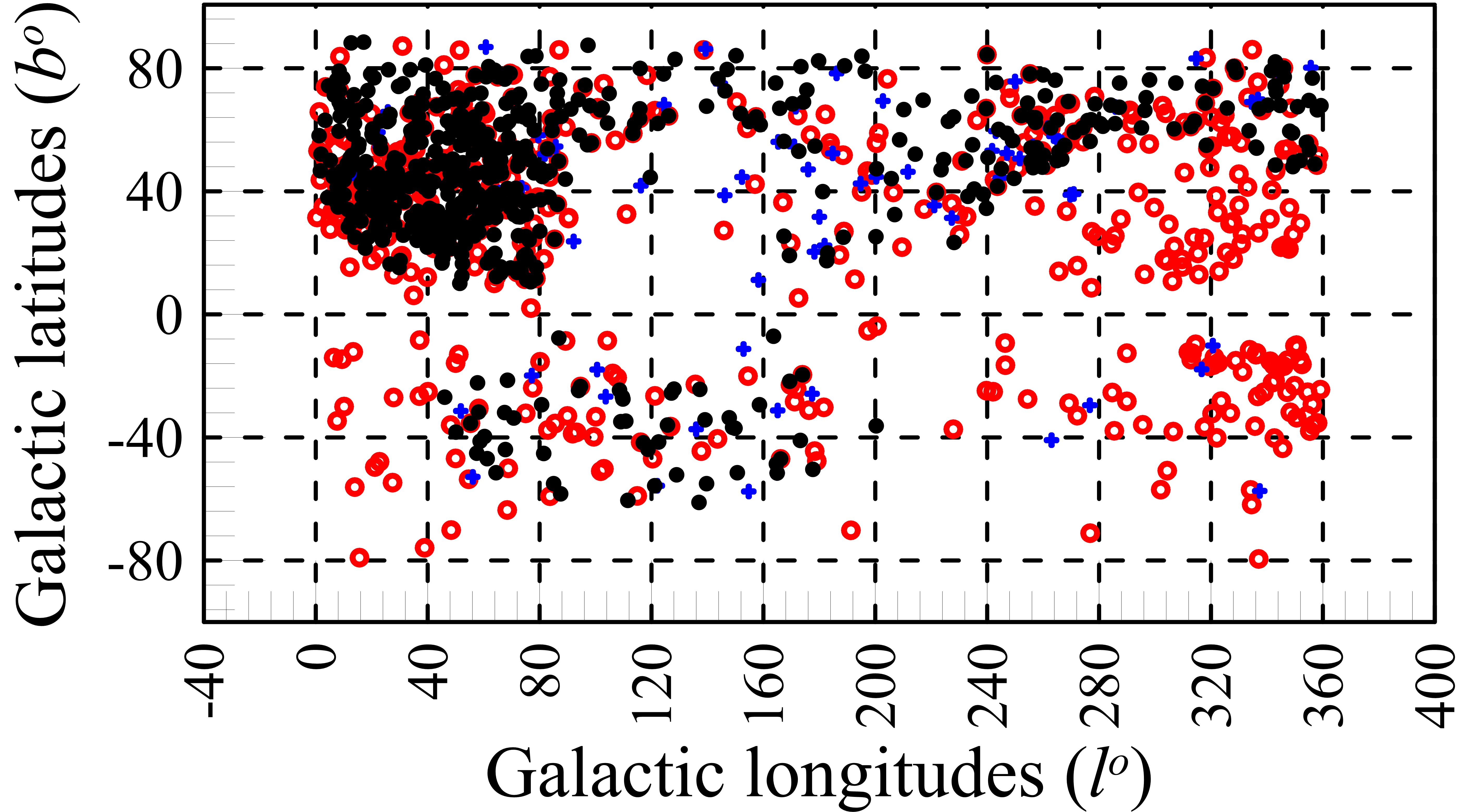}}
\caption{Distribution of HiVels in the galactic coordinate system ($l, b$). Program I: black closed circles (591 stars), Program II: blue closed pluses (87 stars), and Program III: red open circles (519 stars).}
\label{fig1}
\end{figure}

\begin{figure}
\label{fig2}
\centerline{\includegraphics[width=0.75\textwidth,clip=]{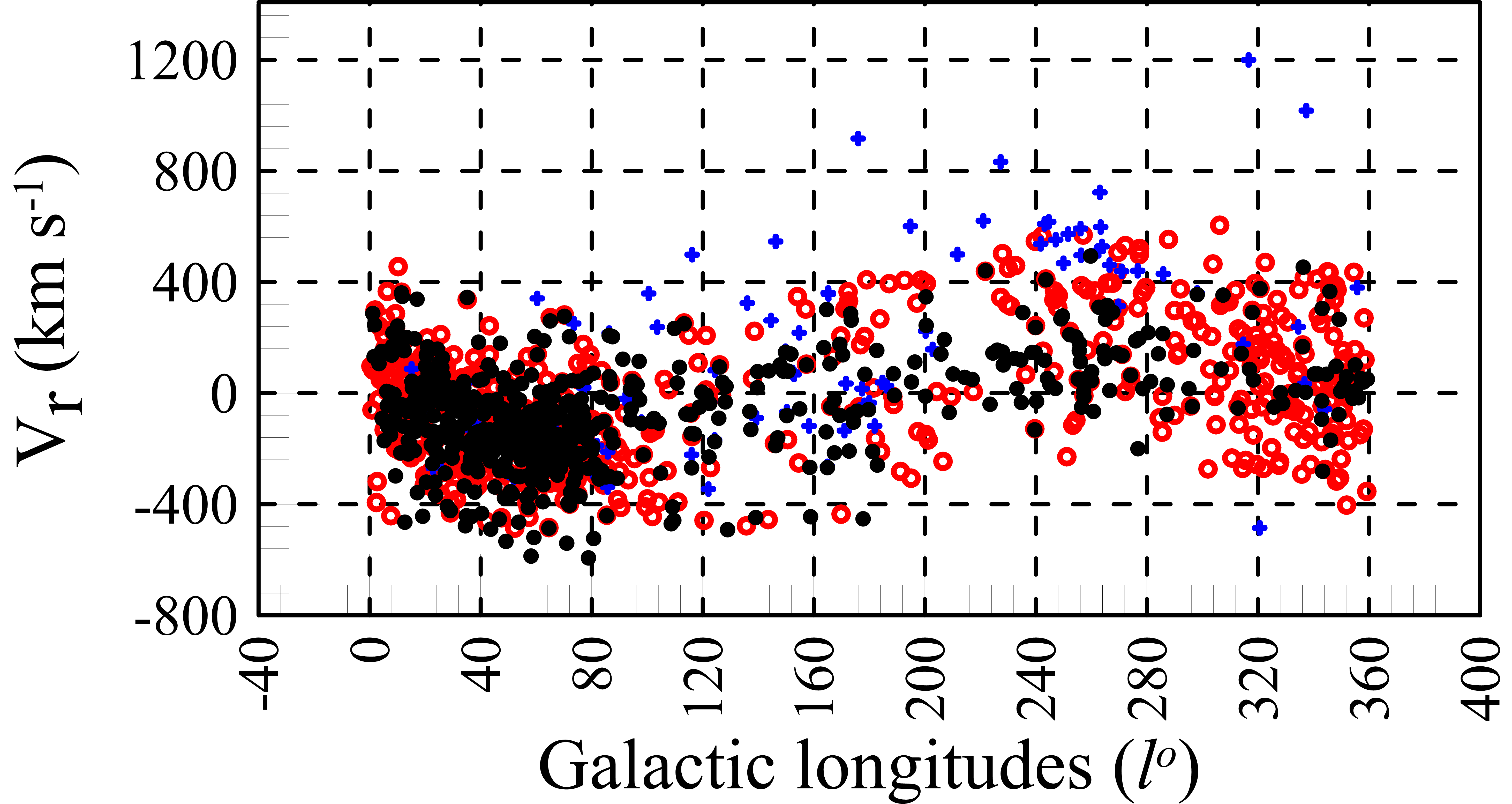}}
\caption{Distribution of ($V_r$; km s$^{-1}$) for three Program HiVels as a function of their galactic longitudes ($l^o$). Program I: black closed circles (591 stars), Program II: blue closed pluses (87 stars), and Program III: red open circles (519 stars).}
\label{fig2}
\end{figure}

\section{VEPs and CP}
\label{sec3}
\subsection{Velocity Ellipsoid Parameters (VEPs)}
A equatorial-galactic transformation matrix based on the SPECFIND v2.0 catalog of radio continuum spectra (see Eq. (14); \cite{2011A&A...536A.102L}) was used to derive the spatial velocity components ($U$, $V$, and $W$; km s$^{-1}$) of HiVels in galactic coordinates with the aid of the calculated space velocity components ($V_x$, $V_y$, and $V_z$; km s$^{-1}$) at distances ($d_i$; pc) from the Sun as given in Eqs. (\ref{Vx}, \ref{Vy}, and \ref{Vz}). Therefore,

\begin{equation}
U= -0.0518807421V_{x} - 0.8722226427V_y -0.4863497200 V_z
\end{equation}
\begin{equation}
V=0.4846922369V_{x}-0.4477920852V_{y}+0.7513692061V_{z},
\end{equation}
\begin{equation}
W=-0.8731447899V_{x}-0.1967483417V_{y}+0.4459913295V_z
\end{equation}
where
\begin{equation}
\label{Vx}
V_x=-4.74d_i\mu_\alpha\cos\delta\sin\alpha-4.74d_i\mu_\delta\sin\delta\cos\alpha +V_r\cos\delta\cos\alpha, \\
\end{equation}
\begin{equation}
\label{Vy}
V_y =+4.74d_i\mu_\alpha\cos\delta\cos\alpha-4.74d_i\mu_\delta\sin\delta\sin\alpha +V_r\cos\delta\sin\alpha, \\
\end{equation}
\begin{equation}
\label{Vz}
V_z=+4.74d_i\mu_\delta\cos\delta+V_r\sin\delta.\\
\end{equation}

In what follows, we estimate the velocities dispersion ($\sigma_1$, $\sigma_2$, and $\sigma_3$; km s$^{-1}$) using the following equations to specify VEPs as described in the literature \citep{2024NewA..11202258E,elsanhouryAN}:

\begin{equation}
\begin{array}{l} {\sigma_1}=\sqrt{2\rho^{\frac{1}{3}}\cos\frac{\phi}{3}-\frac{k_{1}}{3};}\\
{\sigma_2}=\sqrt{-\rho^{\frac{1}{3}}\left\{\cos\frac{\phi}{3}+\sqrt{3}\sin\frac{\phi}{3}\right\}-\frac{k_{1}}{3};}\\{\sigma_3}
=\sqrt{-\rho^{\frac{1}{3}}\left\{\cos\frac{\phi }{3}-\sqrt{3}\sin\frac{\phi}{3}\right\}-\frac{k_{1}}{3}}.{\rm \; \; \;
\; \; \; \; \; \; \; \; \; \; \; \; \; \; \; \; \; \; \; \; \; } \end{array}
\end{equation}

$\rho$ and $\phi$ are calculated as:

\begin{equation}
\rho=\sqrt{-q^{3} },
\end{equation}
\begin{equation}
x=\rho ^{2}-r^{2},
\end{equation}
\begin{equation}
\phi
=\tan ^{-1} \left(\frac{\sqrt{x} }{r} \right).
\end{equation}

The parameters $q$ and $r$ are given by the equations:

\begin{equation}
q=\frac{1}{3} k_{2} -\frac{1}{9} k_{1}^{2} {\rm \; \; \; \; \; \; ;\; \; }r=\frac{1}{6} \left(k_{1} k_{2} -3k_{3} \right)-\frac{1}{27} k_{1}^{3}
\end{equation}

The coefficients $k_1$, $k_2$, and $k_3$  are determined as a function of matrix elements ($\mu_{ij};~\forall i=1,2,3;~\forall j=1,2,3$):
\begin{equation}
\begin{array}{l} {k_{1}=-\left(\mu _{11} +\mu _{22} +\mu _{33} \right),} \\ {k_{2}=\mu _{11} \mu _{22} +\mu _{11} \mu _{33} +\mu _{22} \mu _{33} -\left(\mu _{12}^{2} +\mu _{13}^{2} +\mu _{23}^{2} \right),} \\ {k_{3}=\mu _{12}^{2} \mu _{33} +\mu _{13}^{2} \mu _{22} +\mu _{23}^{2} \mu _{11} -\mu _{11} \mu _{22} \mu _{33} -2\mu _{12} \mu _{13} \mu _{23} .} \end{array}
\end{equation}

since

\begin{equation}
 \begin{array}{l} {\mu _{11}=\frac{1}{N} \sum _{i=1}^{N}U_{i}^{2}  -\left(\overline{U}\right)^{2} ;{\rm \; \; \; \; \; }\mu _{12}=\frac{1}{N} \sum _{i=1}^{N}U_{i} V_{i}  -\overline{U}\; \overline{V};} \\ {\mu _{13}=\frac{1}{N} \sum _{i=1}^{N}U_{i} W_{i}  -\overline{U}\; \overline{W};{\rm \; \; }\mu _{22}=\frac{1}{N} \sum _{i=1}^{N}V_{i}^{2}  -\left(\overline{V}\right)^{2} ;} \\ {\mu _{23}=\frac{1}{N} \sum _{i=1}^{N}V_{i} W_{i}  -\overline{V}\; \overline{W};{\rm \; \; \; }\mu _{33}=\frac{1}{N} \sum _{i=1}^{N}W_{i}^{2}  -\left(\overline{W}\right)^{2} .} \end{array}
 \end{equation}

 The direction cosines ($l_j,m_j,n_j;~\forall{j=1,2,3}$) for the eigenvalue problem ($\lambda_j$), matrix elements ($\mu_{ij}$), and velocities dispersion ($\sigma_j$) [i.e., $\lambda_j=\sigma_j^2; \forall{j=1,2,3}$] where ($\lambda_1>\lambda_2>\lambda_3$), along three axes \citep{2015RMxAA..51..199E}, are mathematically given as follows:

 \begin{equation}
l_j=\left[\mu_{22} \mu_{33} -\sigma_i^2\left(\mu_{22} +\mu_{33} -\sigma_i^2\right)-\mu_{23}^2\right]\big{/}D_j,
\end{equation}
\begin{equation}
m_j=\left[\mu_{23} \mu_{13} -\mu_{12} \mu_{33} +\sigma_{j}^{2} \mu_{12}\right]\big{/}D_j,
\end{equation}
\begin{equation}
n_j=\left[\mu_{12} \mu_{23} -\mu_{13} \mu_{22} +\sigma_{j}^{2} \mu_{13}\right]\big{/}D_j.
\end{equation}
and
\begin{center}
$D_j^2 =\left(\mu_{22}\mu_{33}-\mu_{23}^{2}\right)^{2}+\left(\mu_{23}\mu_{13}-\mu_{12}\mu_{33}\right)^{2}$\\
$+\left(\mu_{12}\mu_{23}-\mu_{13}\mu_{22}\right)^{2}+2[\left(\mu_{22}+\mu_{33}\right)\left(\mu_{23}^{2}-\mu_{22}\mu_{33}\right)$\\
$+\mu_{12}\left(\mu_{23}\mu_{13}-\mu_{12}\mu_{33}\right)+
\mu_{13}\left(\mu_{12}\mu_{23}-\mu_{13}\mu_{22}\right)]\sigma_j^2$\\
$+\left(\mu_{33}^{2}+4\mu_{22}\mu_{33}+\mu_{22}^{2}-2\mu_{23}^{2}+\mu_{12}^{2}+\mu_{13}^{2}\right)\sigma_j^4$\\
$-2\left(\mu_{22}+\mu_{33}\right)\sigma_j^6+\sigma_j^8$.
\end{center}

where ($l_j^2+m_j^2+n_j^2=1$) is an initial test for our code and ($l_2$) is known as vertex longitude \citep{1983Afz....19..505M,2016Ap.....59..246E}.

\subsection{Galactic longitude and latitude parameters}
Let $L_j$ and $B_j$, ($\forall{j=1, 2, 3}$) be the galactic longitude and latitude of the directions, respectively, which correspond to the extreme values of the dispersion, then
\begin{equation}
L_{j}=tan^{-1}\Big(\frac{-m_{j}}{l_{j}}\Big),
\end{equation}
\begin{equation}
B_{j}=sin^{-1}\Big(n_{j}\Big).
\end{equation}

\subsection{Fundamental solar elements}
For Program stars having a space velocities ($\overline{U},~\overline{V},~\overline{W}$), the components of the Sun's velocities are referred to as ($U_{\odot}, V_{\odot},$ and $W_{\odot}$), where ($U_{\odot}= -\overline{U}$), ($V_{\odot}= -\overline{V}$), and ($W_{\odot}= -\overline{W}$). Therefore, the solar elements ($S_{\odot},~l_A,~b_A$) with spatial velocity considered may take the following:

\begin{equation}
S_{\odot}=\sqrt{\overline{U}^2+\overline{V}^2+\overline{W}^2},
\end{equation}
\begin{equation}
l_{A}=tan^{-1}\Bigg(\frac{-\overline{V}}{\overline{U}}\Bigg),
\end{equation}
\begin{equation}
b_{A}=sin^{-1}\Bigg(\frac{-\overline{W}}{S_{\odot}}\Bigg).
\end{equation}

In what follows, we estimate the Sun's local velocity from radial velocities, by letting $X_{\odot}^{\bullet},Y_{\odot}^{\bullet}$ and $Z_{\odot}^{\bullet}$ be the components of the Sun's velocity relative to $x, y,$ and $z$ axes with the origin at the observer. Therefore, $X_{\odot}^{\bullet}=-\overline{V_x}$, $Y_{\odot}^{\bullet}=-\overline{V_y}$, and $Z_{\odot}^{\bullet}=-\overline{V_z}$. From there, we can find the apex of the Sun's way from formulae,

\begin{equation}
\alpha_{A}=tan^{-1}\Bigg(\frac{Y_{\odot}^{\bullet}}{X_{\odot}^{\bullet}}\Bigg),
\end{equation}
\begin{equation}
\delta_{A}=tan^{-1}\Bigg(\frac{Z_{\odot}^{\bullet}}{\sqrt{\left(X_{\odot}^{\bullet}\right)^2+\left(Y_{\odot}^{\bullet}\right)^2}}\Bigg).
\end{equation}
and the Sun's velocity is given by
\begin{equation}
S_{\odot}=\sqrt{\left(X_{\odot}^{\bullet}\right)^2+\left(Y_{\odot}^{\bullet}\right)^2+\left(Z_{\odot}^{\bullet}\right)^2},
\end{equation}

Where $S_{\odot}$ denotes the absolute value of the Sun's velocity relative to the three Program stars under consideration, ($l_A,~b_A$) are galactic longitude and galactic latitude of the solar apex, respectively, and ($\alpha_A,~\delta_A$) is the galactic right ascension and declination with respective manner of the solar apex.
Figure \ref{fig3}, shows the HiVels in three projection velocities (i.e., $U,~V,$ and $W$) with respect to the galactic center. This allows researchers to examine stellar populations in the circumsolar area of the Galaxy. Of great interest is the UV plane, where all stars can be properly classified into multiple families. Nearby stars mostly originate from the disk, which has a distinctive velocity dispersion $\sim$ -500 km s$^{-1}$ along each coordinate.

following the computational algorithm developed by \cite{2015RMxAA..51..199E} to serve and compute the VEPs and the convergent point (CP) of three HiVels Programs, including various kinematic parameters (e.g., spatial velocities, velocities dispersion, direction cosines, solar elements, ... etc.) and the ratios of ($\sigma_2⁄\sigma_1$) and ($\sigma_3⁄\sigma_1$), into which the obtained results of these ratios ranged from 0.75 to 0.86 $\&$ 0.40 to 0.60, respectively.

Stars in the galactic disk have the largest velocity dispersion in the radial direction, which is why it is said that the velocity ellipsoid's longest axis points approximately in the direction of the galactic center. The velocity ellipsoid's longest axis points roughly in the direction of the galactic center because one of its axes is oriented normally to the plane of the galaxy, allowing the other two axes to also lie there. The angle between a group of stars' average $V_r$ vector and the line connecting the Sun to the galactic center is known as the vertex deviation or longitude of the vertex ($l_2$). The kinematic characteristics of the studied star population (e.g., HiVels) can be examined in this way. In the MW's thin disk, the velocity ellipsoid is oriented almost along the radial direction (toward the galactic center) because the vertex's longitude is usually near 0$^{o}$ \citep{2015RMxAA..51..199E,2016Ap.....59..246E,Elsanhoury2020,1981gask.book.....M}. The original numerical results are listed in Table \ref{tab2} and Table \ref{tab3}, including the solar elements.

\begin{figure}
\label{fig3}
\centerline{\includegraphics[width=0.45\textwidth,clip=]{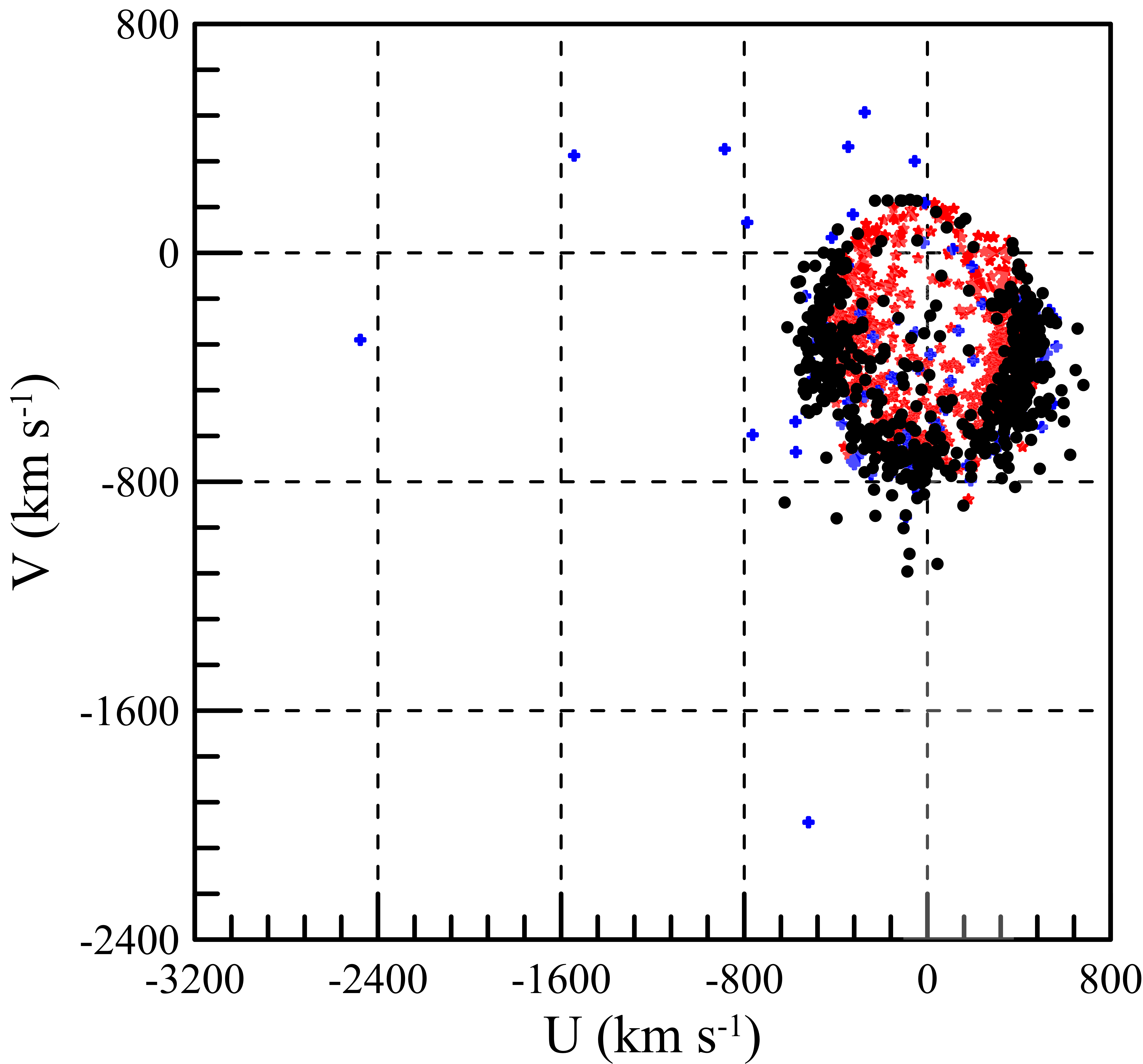}\includegraphics[width=0.45\textwidth,clip=]{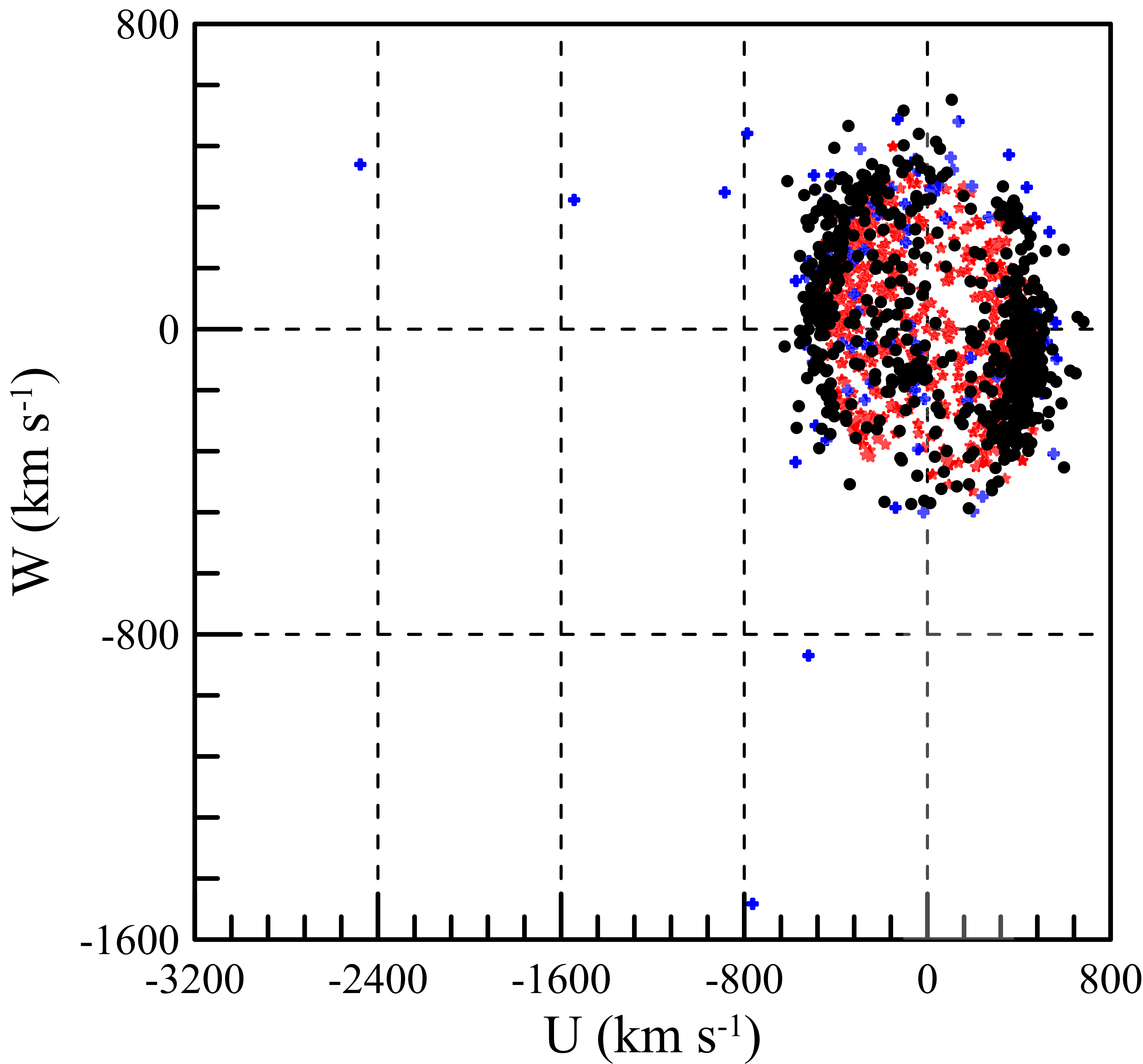}\includegraphics[width=0.45\textwidth,clip=]{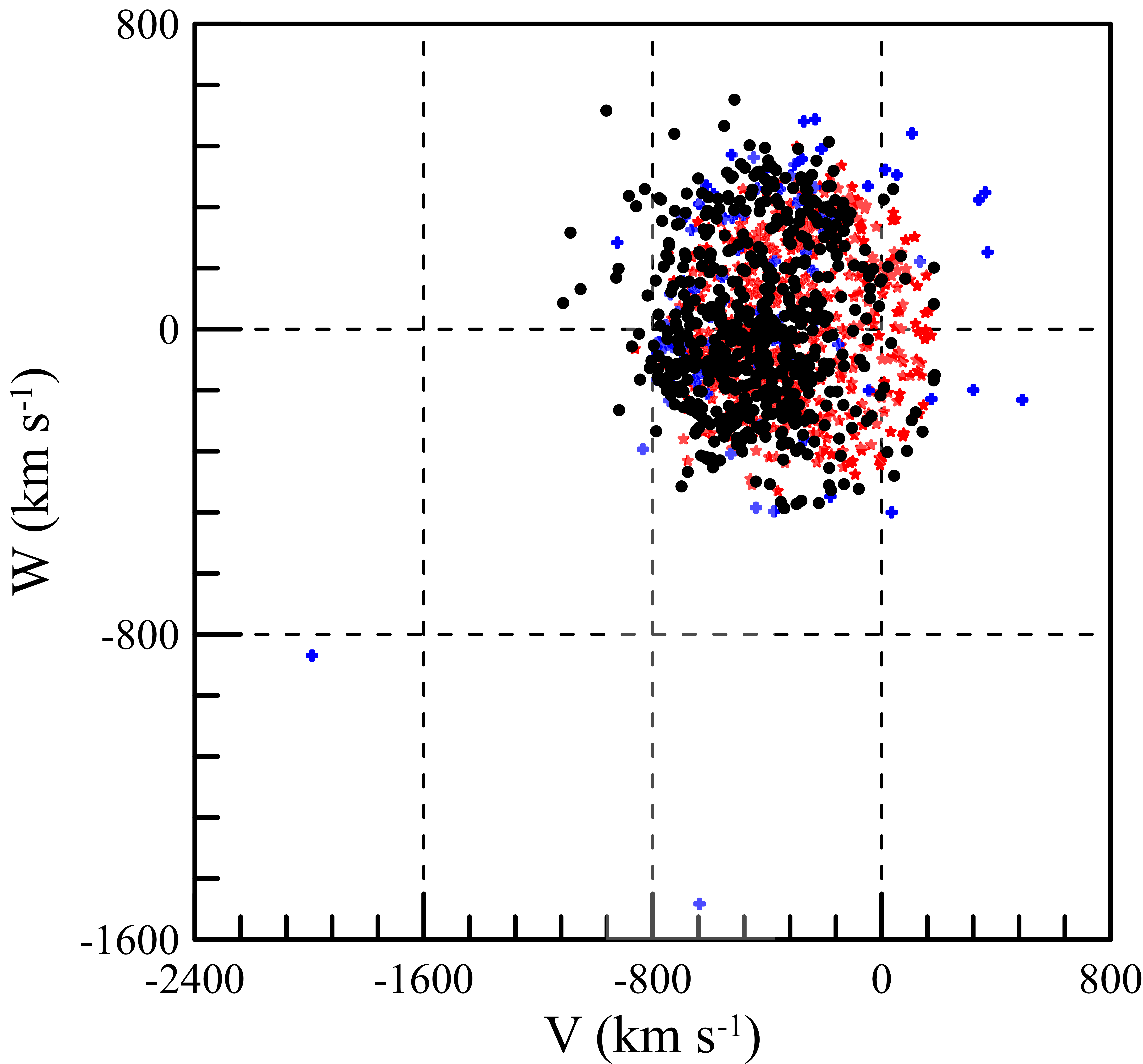}}
\caption{Velocity components distributions $UV$, $UW$, and $VW$ relative to the Galaxy center of three Program HiVels calculated by us. Program I: black closed circles – 591 HiVels, Program II: blue closed pluses – 87 HiVels, and Program III: red open circles – 519 stars.}
\label{fig3}
\end{figure}

\subsection{Convergent Point (CP)}
Usually, stars in a stellar association and/or moving group have a common chemical composition, space velocity, age, and distance which indicates that they are moving across the Galaxy together. The apparent movement of these stars as viewed by the observer, however, varies slightly based on their position in the MW. Their $PM$ vectors indicate the velocity and bearing of the stars' passage across the MW. Typically, in stellar kinematics, the coherent point in the direction of one location on the celestial sphere is well-defined as a vertex, apex, or CP with equatorial coordinates ($A_o$, $D_o$). Several techniques are stated for this purpose like i) the classical CP method \citep{Boss1908}, ii) the individual star apex (AD-diagram), known and developed by \cite{Chupina01,Chupina06}, and iii) the convergent point search method (CPSM) edited by \cite{2012A&A...538A..23G}. In present analysis, we specifically focus on the AD chart method, which relies on the distribution of individual apexes within the equatorial coordinate system and was employed \citep{Elsanhoury2020,2021JApA...42...90E,Elsanhoury2021}. Eqs. \ref{Vx}, \ref{Vy}, and \ref{Vz} of average space velocity vectors are used in this manner to get the equatorial coordinates of the ($A_o$, $D_o$) in the following formulae. The obtained results are listed in the last row of Table \ref{tab2}.

\begin{equation}
\label{eq:ad}
\begin{split}
A_o &= \tan^{-1} \Bigg(\frac{\overline{V_y}}{\overline{V_x}}\Bigg) \\
~~~~\&~~~~
D_o &= \tan^{-1} \Bigg(\frac{\overline{V_z}}{\sqrt {\overline{V_x^2} + \overline{V_y^2}}}\Bigg)
\end{split}
\end{equation}

\begin{table}[t]
\label{tab2}
\tiny
\begin{center}
\caption{VEPs and CPs of the three Program HiVels calculated by us.}
\label{t1}
\begin{tabular}{ll}
\hline\hline
Parameters  & Program I: 591 HiVels \\
 & ($0.10 \le d$(kpc)$~\le 15.40$) \\
\hline
$\overline{V_x}$, $\overline{V_y}$, $\overline{V_z}$ (km s$^{-1}$)&-218.43 $\pm$ 14.76, 139.79 $\pm$ 11.83, -343.81 $\pm$ 18.54\\
$\overline{U}$, $\overline{V}$, $\overline{W}$ (km s$^{-1}$)&56.61 $\pm$ 7.52, -426.79 $\pm$ 20.66, 9.88 $\pm$ 3.14\\
$\lambda_1$, $\lambda_2$, $\lambda_3$ (km s$^{-1}$)&239234, 137669, 38478.70\\
$\sigma_1$, $\sigma_2$, $\sigma_3$ (km s$^{-1}$)&489.12 $\pm$ 22.12, 371.04 $\pm$ 19.26, 196.16 $\pm$ 14.01\\
$\overline{\sigma_o}$ (km s$^{-1}$)&644.50 $\pm$ 15.40\\
($l_1$, $m_1$, $n_1$)$^{o}$&0.2869 $\pm$ 0.002, -0.9578 $\pm$ 0.002, -0.0220 $\pm$ 0.001\\
($l_2$, $m_2$, $n_2$)$^{o}$&-0.9145 $\pm$ 0.003, -0.2806 $\pm$ 0.002, 0.2916 $\pm$ 0.002\\
($l_3$, $m_3$, $n_3$)$^{o}$&0.2854 $\pm$ 0.002, 0.0636 $\pm$ 0.001, 0.9563$\pm$ 0.002\\
$L_j$, $j=1,2,3$&73.33, 162.94, 167.45\\
$B_j$, $j=1,2,3$&-1.26, 16.95, 73.00\\
$S_\odot$ (km s$^{-1}$)&430.641 $\pm$ 20.75\\
($l_A$,~$b_A$)$^{o}$&82.44 $\pm$ 0.11, -1.32 $\pm$ 0.02\\
($\alpha_A$,~$\delta_A$)$^{o}$&-32.62 $\pm$ 0.17, -52.97 $\pm$ 0.86\\
($A_o$,~$D_o$)$^{o}$&147.382 $\pm$ 0.08, -52.973 $\pm$ 0.14\\
\hline
Parameters  & Program II: 87 HiVels \\
 & ($0.30 \le d$(kpc)$~\le 108.64$) \\
\hline
$\overline{V_x}$, $\overline{V_y}$, $\overline{V_z}$ (km s$^{-1}$)&-256.92 $\pm$ 16.03, 276.45 $\pm$ 16.63, -232.17 $\pm$ 15.24\\
$\overline{U}$, $\overline{V}$, $\overline{W}$ (km s$^{-1}$)&114.88 $\pm$ 10.72, -422.77 $\pm$ 20.56, 66.39 $\pm$ 8.15\\
$\lambda_1$, $\lambda_2$, $\lambda_3$ (km s$^{-1}$)&309649, 226100, 108921\\
$\sigma_1$, $\sigma_2$, $\sigma_3$ (km s$^{-1}$)&556.46 $\pm$ 23.59, 475.50 $\pm$ 21.81, 330.03 $\pm$ 18.17\\
$\overline{\sigma_o}$ (km s$^{-1}$)&802.92 $\pm$ 28.34\\
($l_1$, $m_1$, $n_1$)$^{o}$&0.2759 $\pm$ 0.002, 0.9610 $\pm$ 0.002, 0.0181 $\pm$ 0.001\\
($l_2$, $m_2$, $n_2$)$^{o}$&-0.9528 $\pm$ 0.003, 0.2710 $\pm$ 0.002, 0.1368 $\pm$ 0.001\\
($l_3$, $m_3$, $n_3$)$^{o}$&0.1265 $\pm$ 0.001, -0.0550 $\pm$ 0.001, 0.9904 $\pm$ 0.002\\
$L_j$, $j=1,2,3$&-73.98, -164.13, -156.51\\
$B_j$, $j=1,2,3$&1.04, 7.86, 82.07\\
$S_\odot$ (km s$^{-1}$)&443.100 $\pm$ 21.05\\
($l_A$,~$b_A$)$^{o}$&-74.80 $\pm$ 0.12, -8.62 $\pm$ 0.34\\
($\alpha_A$,~$\delta_A$)$^{o}$&-47.10 $\pm$ 0.15, -31.60 $\pm$ 0.18\\
($A_o$,~$D_o$)$^{o}$&132.902 $\pm$ 0.09, -31.600 $\pm$ 0.18\\
\hline
Parameters  & Program III: 519 HiVels \\
& ($0.29 \le d$(kpc)$~\le 16.44$) \\
\hline
$\overline{V_x}$, $\overline{V_y}$, $\overline{V_z}$ (km s$^{-1}$)&-157.41 $\pm$ 12.55, 121.11 $\pm$ 11.00, -230.11 $\pm$ 15.17\\
$\overline{U}$, $\overline{V}$, $\overline{W}$ (km s$^{-1}$)&14.46 $\pm$ 3.80, -303.43 $\pm$ 17.42, 10.98 $\pm$ 3.31\\
$\lambda_1$, $\lambda_2$, $\lambda_3$ (km s$^{-1}$)&143180, 80747.10, 34595.50\\
$\sigma_1$, $\sigma_2$, $\sigma_3$ (km s$^{-1}$)&378.39 $\pm$ 19.45, 284.16 $\pm$ 16.86, 186.00 $\pm$ 13.64\\
$\overline{\sigma_o}$ (km s$^{-1}$)&508.45 $\pm$ 22.55\\
($l_1$, $m_1$, $n_1$)$^{o}$&0.1307 $\pm$ 0.001, -0.9910 $\pm$ 0.002, -0.0275 $\pm$ 0.002\\
($l_2$, $m_2$, $n_2$)$^{o}$&-0.9403 $\pm$ 0.003, -0.1327 $\pm$ 0.001, 0.3134 $\pm$ 0.003\\
($l_3$, $m_3$, $n_3$)$^{o}$&0.3142 $\pm$ 0.003, 0.0151 $\pm$ 0.001, 0.0950 $\pm$ 0.002\\
$L_j$, $j=1,2,3$&82.49, 171.97, 177.24\\
$B_j$, $j=1,2,3$&-1.57, 18.27, 71.67\\
$S_\odot$ (km s$^{-1}$)&303.977 $\pm$ 17.43\\
($l_A$,~$b_A$)$^{o}$&87.27 $\pm$ 0.11, -2.07 $\pm$ 0.03\\
($\alpha_A$,~$\delta_A$)$^{o}$&-37.57 $\pm$ 0.16, -49.20 $\pm$ 0.14 \\
($A_o$,~$D_o$)$^{o}$&142.43 $\pm$ 0.08, -49.21 $\pm$ 0.14\\
\hline\hline
\label{tab2}
\end{tabular}
\end{center}
\end{table}

\begin{table}[t]
\label{tab3}
\small
\begin{center}
\caption{Ratios of velocity dispersions and the solar velocities for our three program HiVels and other components of the disks with different authors.}
\label{t1}
\begin{tabular}{lllll}
\hline\hline
Type  & $S_{\odot}$ (km s$^{-1}$) &    $\sigma_2/\sigma_1$ & $\sigma_3/\sigma_1$ & References\\
\hline
Program I – 591 stars&430.641 $\pm$ 20.75&0.76&0.40&Current Study\\
Program II – 87 stars&443.100 $\pm$ 21.05&0.86&0.60&Current Study\\
Program III – 591 stars&303.977 $\pm$ 17.43&0.75&0.49&Current Study\\
Inner halo ($d\le 15$ kpc)&213.36 $\pm$ 14.61&0.70&0.52&\cite{Nouh..and..Elsanhoury..20}\\
Outer halo ($d=15-20$ kpc)&210.14 $\pm$ 14.50&0.76&0.61&\cite{Nouh..and..Elsanhoury..20}\\
Halo disk&-&0.56&0.56&\cite{2020ApJ...903..131Y}\\
Thin disk&-&0.57&0.46&\cite{2020ApJ...903..131Y}\\
Thick disk&-&0.57&0.52&\cite{2020ApJ...903..131Y}\\
Thin disk&-&0.62&0.62&\cite{Soubiran03}\\
Thick disk&-&0.51&0.51&\cite{Soubiran03}\\
$8.9\ge M_V \ge$8.0&-&0.72 $\pm$ 0.04&0.62 $\pm$ 0.04&\cite{1956AJ.....61..228D}\\
$M_V \ge9.0$&-&0.67 $\pm$ 0.05&0.56 $\pm$ 0.04&\cite{1956AJ.....61..228D}\\
\hline\hline
\label{tab3}
\end{tabular}
\end{center}
\end{table}

\section{Galactic rotation constants}
\label{sec4}
The stars offer a perfect sample for understanding the composition and development of the galactic disk and determining the MW constant parameters. These constants are crucial for understanding how stars, gas, and other objects move within the Galaxy's gravitational potential. Moreover, describes the rotation curve and the motion of objects relative to the galactic center. The differential rotation and local angular velocity (i.e., local rotational features) of the MW Galaxy are described by two fundamental parameters, $A$ and $B$ (i.e., Oort's constants \citep{1927BAN.....4...91O}). They give the dynamic structure of the MW, including its rotation curve and the distribution of matter within it. In addition to helping quantify the shear and turbulent flow of the rotation of the Galaxy, these constants are used to examine how the velocity of stars varies with their distance from the galactic center. According to Oort \citep{1927BAN.....4...91O,1927BAN.....3..275O}, the constants ($A~\&~B$) were then calculated using $V_r$ and $PM$, yielding the values $A \approx$ 19 km s$^{-1}$ kpc$^{-1}$ and $B \approx$ -24 km s$^{-1}$ kpc$^{-1}$. In his study, the relatively smooth rotating curve of the Galaxy was demonstrated, cutting out the hypothesis that it rotates like a rigid body.

Following this, numerous attempts have been made using different tracers to calculate the Oort constants and illustrate the galactic rotation. After reviewing the earlier data, \cite{1986MNRAS.221.1023K} came to the following conclusions: $A = 14.40 \pm 1.20$ km s$^{-1}$ kpc$^{-1}$ and $B = -12.00 \pm 2.80$ km s$^{-1}$ kpc$^{-1}$. To determine that $A = -14.82 \pm 0.84$ km s$^{-1}$ kpc$^{-1}$ and $B = -12.37 \pm 0.64$ km s$^{-1}$ kpc$^{-1}$, \cite{1997MNRAS.291..683F} adopted 220 galactic Cepheids with Hipparcos $PM$. Considering $R_o=7.66 \pm 0.32$ kpc by \cite{1998AJ....115..635M}, therefore, the obtained comparable outcome for $A=15.50 \pm 0.40$ km s$^{-1}$ kpc$^{-1}$ using Cepheids. Recently, \cite{Nouh..and..Elsanhoury..20} obtained a comparable outcome of $A~\&~B$ with 15.60 $\pm$ 1.60 $\&$ -13.90 $\pm$ 1.80 km s$^{-1}$ kpc$^{-1}$, \cite{2021AN....342..989E} computed Oort constants like $A = 14.69 \pm 0.61$ km s$^{-1}$ kpc$^{-1}$ and $B = -16.70 \pm 0.67$ km s$^{-1}$ kpc$^{-1}$, and \cite{2024NewA..11202258E} have $A = 12.91 \pm 0.16$ km s$^{-1}$ kpc$^{-1}$ and $B = -13.16 \pm 0.27$ km s$^{-1}$ kpc$^{-1}$.

Here, we the calculate the Oort constants ($A~\&~B$) considered for three-Program HiVels using the LAMOST and DR3 archives. With an amplitude that rises linearly with distance, we follow the discovery that the heliocentric $V_r$ shows a double sine-wave variation with galactic longitude \citep{1974MNRAS.167..621B}.

\begin{equation}
\label{eq26}
V_r=-2A(R_{gc}-R_o)\sin{l}\cos{b}+K, 
\end{equation}

where ($K$; km s$^{-1}$) is a correction term that can be understood as systematic motions of massive stellar groups, systematic inaccuracies in the $V_r$ caused by motions within stellar atmospheres, gravitational redshift, and erroneous wave-length combinations \citep{1965MNRAS.130..245F}, where $l$ and $b$ stand for the specific star's longitude and latitude, respectively, $R_o = 8.20 \pm 0.10$ kpc \citep{Bland19} is the distance from the Sun to the galactic center, and $R_{gc}$ is the star's radial distance from the galactic center, is the cylindrical radius vector, and it is determined by

\begin{equation}
R_{gc}^{2}=R_o^{2}+d^2-2R_o d\cos{l}.
\end{equation}
Table \ref{tab4} lists our three HiVels Investigations for which we computed the Oort constants $A$ and $B$. Column 3's least-squares fit of Eq. \ref{eq26} yields the first Oort constant ($A;$ km s$^{-1}$ kpc$^{-1}$), while the fourth column's is the second Oort constant ($B;$ km s$^{-1}$ kpc$^{-1}$) calculated using the relation ($\sigma_2⁄\sigma_1$)$^2=-B⁄(A-B)$ \citep{Elsanhoury2020}. The ratio ($\sigma_2⁄\sigma_1$) is calculated using the computational method covered in Section \ref{sec3}. The angular velocity is shown in column 5 ($|A-B|$; km s$^{-1}$ kpc$^{-1}$), and the rotational velocity $V_o$ calculated using the well-known relation $V_o = |A-B|~R_o$, where $R_o = 8.20 \pm 0.10$ kpc and is stated in the last column. Our mean Oort and rotational constants as compared with previous researchers are presented in Table \ref{tab5}, it's clear that Oort constants and galactic rotational parameters were computed for about 304,267 main sequence stars from the Gaia DR1 using data at a typical heliocentric distance of 230 pc \citep{Bovy..17}, a sample of stars within 500 parsecs used the Gaia DR2 data \citep{2019ApJ...872..205L}, using the trigonometric parallaxes and $PMs$ of over 25,000 young stars from the Gaia DR2 dataset \citep{2020AstL...46..370K}, using a sample of halo red giants and the radial and spatial velocities of 1,583 red giant stars collected from the SEGUE-1 and SEGUE-2 surveys \citep{Nouh..and..Elsanhoury..20}, sample of 5,627 $A$-type stars selected from the LAMOST surveys that were located within 0.60 kpc \citep{2021MNRAS.504..199W}, devoted mid to late $M$-type stars \citep{2021AN....342..989E}, for a clean sample of stars (130,665) within 100 pc \citep{2023Univ....9..252G} computed these constants with a maximum likelihood model by using the Gaia Catalog of Nearby Stars (GCNS) \citep{2021A&A...649A...6G}, and for high and low galactic latitudes of $K$ dwarfs \citep{elsanhouryAN}.

\begin{table}[t]
\label{tab4}
\tiny
\begin{center}
\caption{Velocity dispersion and rotation constants for three program HiVels under study.}
\label{t1}
\begin{tabular}{llllll}
\hline\hline
Program   & $\sigma_2/\sigma_1$ &    $A$ & $B$ & $|A-B|$& $V_o$\\
HiVels&  &   km s$^{-1}$ kpc$^{-1}$ & km s$^{-1}$ kpc$^{-1}$ & km s$^{-1}$ kpc$^{-1}$& km s$^{-1}$\\
\hline
I (0.10 $\le d$(kpc) $\le$ 15.40); 591 stars&0.76&15.427 $\pm$ 0.25&-21.095 $\pm$ 0.22&36.522 $\pm$ 6.04&299.48 $\pm$ 17.31\\
II (0.30 $\le d$ (kpc) $\le$ 108.64); 87 stars&0.86&3.883 $\pm$ 0.51&-11.029 $\pm$ 0.30&14.912 $\pm$ 3.86&122.28 $\pm$ 11.06\\
III (0.29 $\le d$(kpc) $\le$ 16.44); 519 stars&0.75&16.510 $\pm$ 0.25&-21.227 $\pm$ 0.22&37.737 $\pm$ 6.14&309.44 $\pm$ 17.59\\
\hline\hline
\label{tab4}
\end{tabular}
\end{center}
\end{table}

\begin{table}[t]
\label{tab5}
\small
\begin{center}
\caption{Velocity dispersion and rotation constants for three program HiVels under study.}
\label{t1}
\begin{tabular}{lllll}
\hline\hline
$A$ & $B$ & $|A-B|$& Methods & References\\
km s$^{-1}$ kpc$^{-1}$ & km s$^{-1}$ kpc$^{-1}$ & km s$^{-1}$ kpc$^{-1}$& $V_r$/$PMs$ &\\
\hline
11.94 $\pm$ 0.29&-17.78 $\pm$ 0.24&25.07 $\pm$ 5.01&$V_r$&Current Study\\
15.30 $\pm$ 0.40&-11.90 $\pm$ 0.40&27.20&$PMs$&\cite{Bovy..17}\\
15.10 $\pm$ 0.10&-13.40 $\pm$ 0.40&28.40&$PMs$&\cite{2019ApJ...872..205L}\\
15.73 $\pm$ 0.32&-12.67 $\pm$ 0.34&28.40&$PMs$&\cite{2020AstL...46..370K}\\
15.60 $\pm$ 1.60&-13.90 $\pm$ 1.80&29.50 $\pm$ 0.20&$V_r$&\cite{Nouh..and..Elsanhoury..20}\\
16.31 $\pm$ 0.89&-11.99 $\pm$ 0.79&28.30&$PMs$&\cite{2021MNRAS.504..199W}\\
14.69 $\pm$ 0.61&-16.70 $\pm$ 0.67&31.39&$V_r$&\cite{2021AN....342..989E}\\
15.60 $\pm$ 1.60&-15.80 $\pm$ 1.70&31.40 $\pm$ 2.30&$V_r$&\cite{2023Univ....9..252G}\\
12.91 $\pm$ 0.16&-13.16 $\pm$ 0.27&26.06&$V_r$&\cite{elsanhouryAN}\\
\hline\hline
\label{tab5}
\end{tabular}
\end{center}
\end{table}

\section{Discussion and conclusion}
\label{sec5}
The velocity distribution ($U$, $V$, $W$) of stars in three spatial directions ($\sigma_1,~\sigma_2,~\sigma_3$) is known as the velocity ellipsoid in galactic kinematics. The ellipsoid's shape reflects the anisotropy in the star motion, and its primary axes relate to the directions with the largest and smallest velocity dispersion. Velocity ellipsoids vary among star populations. For instance, thick disk and halo stars exhibit more isotropic ellipsoids and bigger velocity dispersions than thin disk stars, which have comparatively small dispersions. In this study, we calculated the kinematical parameters, convergent point, and the Oort constants $A$ and $B$ of three Programs (i.e., 591, 87, and 519 stars) of high velocity stars located ($0.10~\geq~d (kpc)~\geq~109$). The following summarizes the main findings of the current studies:

$\bullet$~~We retrieved high velocities in both mean spatial velocities ($U,~V,~W$, and $V_{space}$; km s$^{-1}$), Program I (56.61 $\pm$ 7.52, -426.79 $\pm$ 20.66, 9.88 $\pm$ 3.14, and 430.641 $\pm$ 20.75), Program II (114.88 $\pm$ 10.72, -422.77 $\pm$ 20.56, 66.39 $\pm$ 8.15, and 443.100 $\pm$ 21.05), and Program III (14.46 $\pm$ 3.80, -303.43 $\pm$ 17.42, 10.98 $\pm$ 3.31, an 303.977 $\pm$ 17.43) and the mean velocity dispersion ($\sigma_1,~\sigma_2,~\sigma_3$, and $\sigma_o$; km s$^{-1}$), Program I (489.12 $\pm$ 22.12, 371.04 $\pm$ 19.26, 196.16 $\pm$ 14.01, and 644.50 $\pm$ 15.40), Program II (556.46 $\pm$ 23.59, 475.50 $\pm$ 21.81, 330.03 $\pm$ 18.17, and 802.92 $\pm$ 28.34), and Program III (378.39 $\pm$ 19.45, 284.16 $\pm$ 16.86, 186.00 $\pm$ 13.64, and 508.45 $\pm$ 22.55).

$\bullet$~~Our obtained results of the $l_2$ are: -0$^o$.9145 $\pm$ 0$^o$.003 (Program I), -0$^o$.9528 $\pm$ 0$^o$.003 (Program II), and -0.9403 $\pm$ 0$^o$.003 (Program III), this obtained results are in line with that of \cite{1981gask.book.....M} and many authors, e.g., $l_{2}=-0^\circ.3454$, $-0^\circ.6735$, and $-0^\circ.0264$ \citep{2024NewA..11202258E} and $-0^o.87$ $\&$ $-0^o.91$ \citep{Elsanhoury18}.

$\bullet$~~We have determined the convergent points ($A_o,~D_o$), Program I: 147$^o$.382 $\pm$ 0$^o$.08, -52$^o$.973 $\pm$ 0$^o$.14, Program II: 132$^o$.902 $\pm$ 0$^o$.09, -31$^o$.600 $\pm$ 0$^o$.18, and Program III: 142$^o$.43 $\pm$ 0$^o$.08, -49$^o$.21 $\pm$ 0$^o$.14.

$\bullet$~~The mean values of the Oort's constants, $A = 11.94 \pm 0.29$ and $B = -17.78 \pm 0.24$ km s$^{-1}$ kpc$^{-1}$ as listed in Table \ref{tab5}. Therefore, the mean angular velocity $|A-B| = 25.07 \pm 5.01$ km s$^{-1}$ kpc$^{-1}$ and the rotational velocity $V_o = 243.72 \pm 15.61$ km s$^{-1}$.

$\bullet$~~In regions where stars are to be expected to be quite a way distant on average like our consideration of HiVels. The measured values of $A$ and $B$ may contain significant inaccuracies due to both long distances \citep{1990MNRAS.244..247L} and the choice of kinematic model ($V_r$ or $PM$) \citep{1987AJ.....94..409H} as Table \ref{tab4} makes clear with Program II. \cite{1990MNRAS.244..247L} reported that this is due to two effects. First, the proper motion along declination $\mu_{\delta}$ nearly equal to the ratio of ($V_y⁄d$), i.e., $\mu_{\delta} \sim$ ($V_y⁄d$) and $A\sim \mu_{\delta}$, thus if $d$ increases, the value of $A$ will decrease, as clearly seen in Program II of Table \ref{tab4}. But $V_y$ also dependent on ($\cos{\alpha}$) and is dependent on distance $d$ (see Eq. \ref{Vy}). Therefore, these two effects should affect both $A$ and $B$ in a roughly equal manner \citep{1990MNRAS.244..247L}.

\acknowledgements
We sincerely thank the anonymous referee for their valuable suggestions, which have significantly improved the quality of this paper. This work presents results from the European Space Agency space mission Gaia. Gaia data are being processed by the Gaia Data Processing and Analysis Consortium (DPAC). Funding for the DPAC is provided by national institutions, in particular, the institutions participating in the Gaia MultiLateral Agreement (MLA). The Gaia mission website is \url{https://www.cosmos.esa.int/gaia}. The Gaia archive website is \url{https://archives.esac.esa.int/gaia}. The author extend their appreciation to the Deanship of Scientific Research at Northern Border University, Arar, KSA for funding this research work through the project number "NBU-FFR-2025-237-03”.

Guoshoujing Telescope (Large Sky Area Multi-Object Fiber Spectroscopic Telescope LAMOST) is a National Major Scientific Project built by the Chinese Academy of Sciences. Funding for the project has been provided by the National Development and Reform Commission. LAMOST is operated and managed by the National Astronomical Observatories, Chinese Academy of Sciences.\\

{\bf Data availability}: 
We have used the different data sets for high velocity stars, which are publicly available at the following links:
\begin{itemize}  
\setlength{\itemsep}{0pt plus 1pt}
\item \url{https://vizier.cds.unistra.fr/viz-bin/VizieR?-source=J/ApJS/252/3}
\item \url{https://vizier.cds.unistra.fr/viz-bin/VizieR?-source=J/AJ/166/12}
\item \url{https://vizier.cds.unistra.fr/viz-bin/VizieR?-source=J/AJ/167/76}
\item \url{https://vizier.cds.unistra.fr/viz-bin/VizieR-3?-source=I/355/gaiadr3&-out.max=50&-out.form=HTML%20Table&-out.add=_r&-out.add=_RAJ,_DEJ&-sort=_r&-oc.form=sexa}

\end{itemize}

\bibliography{Final_version}

\clearpage
\appendix{Special Signs}
You may wish to use special signs. A large number of the latter are listed
in the {\em {\LaTeX} User's Guide \& Reference Manual\/} by Leslie Lamport,
pp.\,44\,ff. We have created further symbols for math mode, and some special
Slovak characters (used e.g. in names), which cannot be simply produced by
{\LaTeX}.
\begin{table}[htbp]
\caption{Special signs}
\label{tspsig}
\begin{center}
\footnotesize
\renewcommand{\arraystretch}{1.25}
\begin{tabular}{lll@{\hspace{0.5cm}}lll}
\hline\hline
Input & Explanation & Output & Input & Explanation & Output\\
\hline
\verb|\la|     & less or approx       & $\la$     &
  \verb|\ga|     & greater or approx    & $\ga$\\
\verb|\getsto| & gets over to         & $\getsto$ &
  \verb|\cor|    & corresponds to       & $\cor$\\
\verb|\lid|    & less or equal        & $\lid$    &
  \verb|\gid|    & greater or equal     & $\gid$\\
\verb|\sol|    & similar over less    & $\sol$    &
  \verb|\sog|    & similar over greater & $\sog$\\
\verb|\lse|    & less over simeq      & $\lse$    &
  \verb|\gse|    & greater over simeq   & $\gse$\\
\verb|\grole|  & greater over less    & $\grole$  &
  \verb|\leogr|  & less over greater    & $\leogr$\\
\verb|\loa|    & less over approx     & $\loa$    &
  \verb|\goa|    & greater over approx  & $\goa$\\
\hline
  \verb|\degr|     & degree                & $\degr$\\
\verb|\diameter| & diameter              & \diameter  &
  \verb|\sq|       & square                & \squareforqed\\
\verb|\fd|       & fraction of day       & \fd        &
  \verb|\fh|       & fraction of hour      & \fh\\
\verb|\fm|       & fraction of minute    & \fm        &
  \verb|\fs|       & fraction of second    & \fs\\
\verb|\fdg|      & fraction of degree    & \fdg       &
  \verb|\fp|       & fraction of period    & \fp\\
\verb|\farcs|    & fraction of arcsecond & \farcs     &
  \verb|\farcm|    & fraction of arcmin    & \farcm\\
\verb|\arcsec|   & arcsecond             & \arcsec    &
  \verb|\arcmin|   & arcminute             & \arcmin\\
\verb|\angstrom| & angstr\"{o}m          & \angstrom    &
                   &                       &       \\
\hline
\verb|\softL|    & Slovak L with caron   & \softL     &
  \verb|\softl|    & Slovak l with caron   & \softl\\
\verb|\softd|    & Slovak d with caron   & \softd     &
  \verb|\softt|    & Slovak t with caron   & \softt\\
\hline\hline
\end{tabular}
\renewcommand{\arraystretch}{1}
\end{center}
\end{table}

\end{document}